UDC 533.9.072

# INTEGRAL EQUATION FOR SOURCE OF IONIZATION IN HOLLOW CATHODE

**Vladimir V. Gorin**

*National Taras Shevchenko University, Radiophysical Faculty*
*Prospect Glushkova 2/5, 03022 Kyiv, Ukraine,*
vgorin@univ.kiev.ua

**Abstract.** To simulate the influence of pendulum effect in hollow cathode configuration on a current-voltage characteristic it is proposed to consider a plane glow discharge in one-dimensional space configuration with two plane parallel cathodes and one plane anode in the middle. A pendulum effect could be switched on or off, with consideration for the anode to be transparent or not for fast ionizing electrons. On the basis of a stationary kinetic equation for ionizing electrons an integral equation for a source of ionization is derived. This equation can be used as a component in self-consistent problem for field equations to calculate a current-voltage characteristic and to show how it changes in both cases – in presence or absence of anode transparency.

**Glossary of terms for physical values**

$x = X(t, v_0), \quad v = V(t, v_0)$ — electron coordinate and velocity

$v_0 = V_0(x, v)$ — initial electron velocity on a cathode

$t = T(x, v)$ — time of electron motion from a cathode to phase point $x, v$

$w = \dfrac{m_e v^2}{2e}$ — electron kinetic energy in ev

$\varphi(x), \quad E = -d\varphi/dx$ — electric field potential and electric field strength

$\varepsilon(x, v) = -\varphi(x) + \dfrac{m_e v^2}{2e}$ — total electron mechanical energy in ev

$U = \varphi(0), \quad x_c$ — cathode-anode voltage and width of cathode-anode gap

$\varepsilon_{ion}, \quad \varepsilon_{ex}$ — ionization and excitation energy of gas molecule in ev

$\sigma_{ion}(w), \quad \sigma_{ex}(w)$ — cross-sections of electron impact ionization and excitation of gas molecule

$N_a$ — gas density

$L(w) = N_a(\varepsilon_{ion}\sigma_{ion}(w) + \varepsilon_{ex}\sigma_{ex}(w))$ — average force of resistance to electron motion, which is stipulated by energy losses for ionization and excitation of gas molecules

$\left(|v|\dfrac{dL}{dw}\right)(v) \equiv |v|\dfrac{dL}{dw}(w), \quad w = \dfrac{m_e v^2}{2e}$ — brief designations in formulas

$f_e(x, v)$ — distribution function of ionizing electrons – average number of electrons per volume unit and unit of velocity interval

$i_e(x) = \displaystyle\int_{-\infty}^{\infty} v f_e(x, v) dv, \quad J_e(x) = -e i_e(x)$ — electron flow and current densities

$s(x)$ — number of ionization acts per volume unit and time unit

$\delta(v)$ — Dirac delta-function

$J_{ec}$ — electron current density in a cathode

$S(x) \equiv s(x)e/J_{ec}$ — number of ionization acts per unit of electron path, which corresponds to one cathode electron

$t_n = t_n(v_0), \quad n = 1, 2, \ldots t_1 < t_2 < \ldots$     instant times of electron reverse

$x_n(v_0) \equiv X(t_n(v_0), v_0)$     coordinates of reverse points

$E_n(v_0) \equiv E(x_n(v_0))$     values of electric field in the reverse points

$K(x, v_0)$     number of roots of equation $X(t, v_0) = x$ with regard to variable $t$

$t_k = t_k(x, v_0)$     instant times, in which electron reaches coordinate $x$: $X(t_k, v_0) = x, \quad 0 < x < x_c$

$v_k(x, v_0) = V(t_k, v_0)$     values of velocity, which correspond to these instant times

$\theta(z): \quad \theta(z) = 0, z \leq 0; \quad \theta(z) = 1, z > 0.$     Heaviside function

$\alpha$     first Townsend coefficient of ionization

Upper point designates time derivative along phase trajectory of electron.

## 1. Problem statement and description of the model

Plane *simple glow discharge* (SGD), by ignoring technical bounds (as did Engel and Shtenbek [1]), could be considered as one-dimensional. In this classical theory of cathode dark space it was consumed small free path for electrons in comparison with a scale of spatial electric field variation. Under this condition the inertia of electron could be neglected and the electron energy in every space point could be considered as a function of electric field in this point only. As a sequence, the local energy dependence was the Townsend's local dependence of ionization capability of electron on electric field: $\alpha = \alpha(E(x))$.

Our aim is to build simple analogue of one-dimensional model for *hollow cathode discharge* (HCD). Under conditions of *hollow cathode effect* (HCE) – multiple exceeding of electric current density in comparison with that was in SGD at the same voltage – it is not possible to neglect the electron inertia because electron oscillations inside a device become significant feature. Here electron energy has non-local dependence on electric field, therefore ionization capability of electron has non-local dependence on electric field also. Thus, the model of a source of ionization in HCD, in distinct of classical one, must be non-local.

We consider one-dimensional plane glow discharge with two plane parallel cathodes and one plane anode in the middle (see fig. 1). We can assume the anode to be a system of plane conducting meshes, which catch slow electrons and mainly do not prevent fast electrons to pass through. Also we assume a discharge to be uniform along two coordinate axes, which are parallel to cathode and anode plates. Here, we neglect elastic electron scattering (it was shown [2], that at energies about 200 – 550 ev cross-sections of elastic processes are several times lower than inelastic ones), as well as ion-electron recombination processes, and we consider electron motion only in direction, which is normal to cathode and anode plates. We regard that electron accelerates with electric field and slows down being involved into ionization and excitation of neutral gas molecules. The influence of these inelastic processes we describe with effective "force of friction".

## 2. Phase space and phase trajectories

Consider properties of phase space (see fig. 2)

$$x, v: \quad -x_c < x < +x_c, \quad -\infty < v < +\infty, \quad \varepsilon(x, v) > \varepsilon_{ion}$$

for a motion of ionizing electron in plane hollow cathode. It is a symmetrical doubly-connected region with exclusion of low mechanical energy region in the middle.

Let $x = X(t, v_0), \quad v = V(t, v_0)$ be the solutions of a Cauchy problem for a system of two ordinary differential equations

$$\begin{cases} \dot{x} = v, \\ \dot{v} = -\dfrac{e}{m_e}\left(E(x) + \operatorname{sgn}(v) L\left(\dfrac{m_e v^2}{2e}\right)\right). \end{cases} \quad (1)$$

$$x\big|_{t=0} = \pm x_c, \quad v\big|_{t=0} = v_0.$$

for positive time $t \geq 0$ with initial point of trajectory in $x = +x_c, v = v_0, v_0 < 0$ or in $x = -x_c, v = v_0, v_0 > 0$. Every point of phase space we can consider such that belongs to some trajectory. Moreover nothing prevents us from considering the trajectory of electron, which is generated inside a volume of the device at $-x_c < x < +x_c$ (secondary electron), to be a piece of trajectory of some cathode electron having fitting value of initial velocity $v_0$ in the cathode, such that a trajectory passes through a point of generation of secondary electron. Points $(x,v)$, $(-x,-v)$ belong to symmetrical trajectories, one of which starts at $x = +x_c, v = v_0$, another - at $x = -x_c, v = -v_0$. As the system of order differential equations (1) is resolved with respect to its derivatives, right sides of the equations are defined everywhere, so trajectories do not cross each other and fill over all phase space.

Let
$$\begin{cases} t = T(x,v), \\ v_0 = V_0(x,v). \end{cases} \tag{2}$$

be a solution of equation system
$$\begin{cases} x = X(t,v_0), \\ v = V(t,v_0). \end{cases} \tag{3}$$

with respect to variables $t, v_0$. Functions (2), (3) have next symmetries of parity:
$$X(t,-v_0) = -X(t,v_0), \quad V(t,-v_0) = -V(t,v_0), \quad T(-x,-v) = T(x,v), \quad V_0(-x,-v) = -V_0(x,v). \tag{4}$$

Couple of numbers $(t,v_0)$ defines phase point $(x,v)$ identically, except for a case of a separatrix trajectory $v_0 = 0$; in this case we can exclude points $(x,v) = (+x_c, 0)$, $(x,v) = (-x_c, 0)$ from a set of initial points for the system (1) by prolonging the separatrix trajectory back to crossing it with another cathode plate (see fig. 2). Negative value of $v_0$ corresponds to trajectory, which starts at a right-hand side cathode plate $x = +x_c$, positive value of $v_0$ corresponds to trajectory, which starts at a left-hand side cathode plate $x = -x_c$.

Consider differential properties of functions (2), which are useful below. From relations
$$\begin{cases} dt = \frac{\partial T}{\partial x} dx + \frac{\partial T}{\partial v} dv, \\ dv_0 = \frac{\partial V_0}{\partial x} dx + \frac{\partial V_0}{\partial v} dv. \end{cases} \qquad \begin{cases} dx = \dot{x}\, dt + \frac{\partial X}{\partial v_0} dv_0, \\ dv = \dot{v}\, dt + \frac{\partial V}{\partial v_0} dv_0. \end{cases}$$

it follows
$$\begin{bmatrix} \dot{x} & \frac{\partial X}{\partial v_0} \\ \dot{v} & \frac{\partial V}{\partial v_0} \end{bmatrix} \cdot \begin{bmatrix} \frac{\partial T}{\partial x} & \frac{\partial T}{\partial v} \\ \frac{\partial V_0}{\partial x} & \frac{\partial V_0}{\partial v} \end{bmatrix} = \begin{bmatrix} \frac{\partial T}{\partial x} & \frac{\partial T}{\partial v} \\ \frac{\partial V_0}{\partial x} & \frac{\partial V_0}{\partial v} \end{bmatrix} \cdot \begin{bmatrix} \dot{x} & \frac{\partial X}{\partial v_0} \\ \dot{v} & \frac{\partial V}{\partial v_0} \end{bmatrix} = \begin{bmatrix} 1 & 0 \\ 0 & 1 \end{bmatrix}. \tag{5}$$

In particular, it contains relations:
$$\dot{x}\frac{\partial V_0}{\partial x} + \dot{v}\frac{\partial V_0}{\partial v} = 0, \tag{6}$$

$$\dot{x}\frac{\partial T}{\partial x} + \dot{v}\frac{\partial T}{\partial v} = 1. \tag{7}$$

From relations (5), with account of equations (1), through elementary transforms one can obtain expressions (see appendix 1):
$$\frac{\partial V_0}{\partial v}(x,v) = \frac{v}{V_0(x,v)} \exp\left( \int_0^{T(x,v)} \left( |v| \frac{dL}{dw} \right)(V(t',V_0(x,v))) dt' \right). \tag{8}$$

$$\frac{\partial V_0}{\partial x}(x,v) = \frac{-\dot{v}}{v}\frac{\partial V_0}{\partial v} = \frac{-\dot{v}(x,v)}{V_0(x,v)}\exp\left(\int_0^{T(x,v)}\left(|v|\frac{dL}{dw}\right)(V(t',V_0(x,v)))dt'\right). \tag{9}$$

### 3. Distribution function

We do not consider the distribution of all electrons together – slow and fast. Slow electrons assumed to obey hydrodynamics of fluid and do not concern to a source of ionization. So, a kinetic equation for fast (ionizing) electrons only would not have the Boltzmann or the Fokker-Plank collision terms, which describe processes of scattering. We neglect angular scattering and include energy losses in ionization and excitation processes only - through an effective force of friction.

The equation for stationary electron distribution function $f_e(x,v)$ should correspond to dynamical equations (1), also it should include a possibility of generation of secondary electron, which arises due to electron impact ionization with negligibly small velocity. So, the distribution function should obey the stationary kinetic equation with a source:

$$\dot{x}\frac{\partial f_e}{\partial x} + \frac{\partial}{\partial v}(\dot{v}(x,v)f_e) = s(x)\delta(v). \tag{10}$$

Here functions $\dot{x},\dot{v}$ are defined with the equations (1). Besides, we should take into account that electron flow density $i_e(x) = \int_{-\infty}^{\infty} v\, f_e(x,v)dv$ obeys a continuity relation $\frac{di_e}{dx} = s(x)$, where $s(x)$ is an ionization source density - a number of ionization acts per volume unit and time unit. Substituting the relations (1) and taking partial derivatives, let us rewrite (10) in the form:

$$v\frac{\partial f_e}{\partial x} - \frac{e}{m_e}\left(E(x) + \text{sgn}(v)L\left(\frac{m_e v^2}{2e}\right)\right)\frac{\partial f_e}{\partial v} - |v|\frac{dL}{dw}f_e = s(x)\delta(v). \tag{11}$$

It is a quasi-linear non-uniform 1-st order partial differential equation [3]. A solution of this equation has a form (see appendix 2):

$$f_e(x,v) = f_e(-x_c\,\text{sgn}(V_0(x,v)),V_0(x,v))\exp\left(\int_0^{T(x,v)} dt'\left(|v|\frac{dL}{dw}\right)(V(t',V_0(x,v)))\right) + \\ + \int_0^{T(x,v)} dt''s(X(t'',V_0(x,v)))\delta(V(t'',V_0(x,v)))\exp\left(\int_{t''}^{T(x,v)} dt'\left(|v|\frac{dL}{dw}\right)(V(t',V_0(x,v)))\right). \tag{12}$$

In another way, from its physical origin, electron source density (it is also ionization source density) is defined with expression

$$s(x) = \int_{-\infty}^{+\infty} dv\, N_a|v|\sigma_{ion}\left(\frac{m_e v^2}{2e}\right)f_e(x,v). \tag{13}$$

So, the combination of (12) and (13), leads to a 2-nd kind Fredholm integral equation with a singular kernel.

Let $t_n = t_n(v_0)$, $n = 1,2,...$ $t_1 < t_2 < ...$ be roots of equation $V(t,v_0) = 0$ (namely, they are instant times, in which cathode electron, having the initial velocity $v_0$, "turns back" and then moves in opposite direction), $x_n(v_0) \equiv X(t_n(v_0),v_0)$ are coordinates $(-x_c < x_n \leq x_c)$ of reverse points and $E_n(v_0) \equiv E(x_n(v_0))$ are the values of electric field strength in these points (see fig. 3).

From equations (1), taking into account the property $L(w) = 0$ at $w \leq \varepsilon_{ex}$, for reverse points we obtain:

$$\dot{V}(t_n,v_0) = -\frac{e}{m_e}E_n, \tag{14}$$

$$\delta(t'' - t_n(v_0)) = \delta(V(t'', v_0)) \left| \frac{dV}{dt}(t'', v_0) \right| = \delta(V(t'', v_0)) |\dot{V}(t_n, v_0)| = \delta(V(t'', v_0)) \frac{e}{m_e} |E_n(v_0)|,$$

$$\delta(V(t'', V_0(x, v))) = \frac{m_e}{e |E_n(V_0(x, v))|} \delta(t'' - t_n(V_0(x, v))).$$

In substitution of this expression second term in (12) can be integrated and the expression gets a form:

$$f_e(x, v) = f_e(-x_c \operatorname{sgn}(V_0(x, v)), V_0(x, v)) \exp\left( \int_0^{T(x,v)} dt' \left( |v| \frac{dL}{dw} \right) (V(t', V_0(x, v))) \right) +$$
$$+ \frac{m_e}{e} \sum_{n: 0 < t_n(V_0(x,v)) < T(x,v)} \frac{s(x_n(V_0(x, v)))}{|E_n(V_0(x, v))|} \exp\left( \int_{t_n(V_0(x,v))}^{T(x,v)} dt' \left( |v| \frac{dL}{dw} \right) (V(t', V_0(x, v))) \right). \quad (15)$$

Here the first term describes a component of distribution function, which is formed by cathode electron ionization only, the second term describes contribution of secondary electron ionization.

### 4. Integral equation for source of ionization

From the couple of equations (13), (15) one can exclude the source, - then one can obtain an integral equation for distribution function. But we commit quite on the contrary: we exclude proper distribution function $f_e(x, v)$:

$$s(x) = \int_{-\infty}^{+\infty} dv \, N_a |v| \sigma_{ion}\left( \frac{m_e v^2}{2e} \right) \times$$
$$\times \left( f_e(-x_c \operatorname{sgn}(V_0(x, v)), V_0(x, v)) \exp\left( \int_0^{T(x,v)} dt' \left( |v| \frac{dL}{dw} \right) (V(t', V_0(x, v))) \right) + \right. \quad (16)$$
$$\left. + \frac{m_e}{e} \sum_{n: 0 < t_n(V_0(x,v)) < T(x,v)} \frac{s(x_n(V_0(x, v)))}{|E_n(V_0(x, v))|} \exp\left( \int_{t_n(V_0(x,v))}^{T(x,v)} dt' \left( |v| \frac{dL}{dw} \right) (V(t', V_0(x, v))) \right) \right).$$

As distinct from integral equation for distribution function $f_e(x, v)$ we obtain an integral equation for function $s(x)$ of not two, but single variable, $x$.

To simplify the kernel structure we transform the integration variable $v$ of velocity in point $x$ to initial velocity $v_0$ in the cathode according to the relation $v_0 = V_0(x, v)$, which now we have to solve with the regard to a variable $v$ at constant $x$. As a phase trajectory can be spiral, this equation may have several of roots $v = v_k(x, v_0)$, $k = 1, 2, ..., K(x, v_0)$ or no roots – at some value of $x, v_0$. Let $t_k = t_k(x, v_0) = T(x, v_k)$, $k = 1, 2, ..., K(x, v_0)$ be instant times when electron is at a coordinate $x$: $X(t_k, v_0) = x$, $w_k(x, v_0) \equiv \frac{m_e v_k^2(x, v_0)}{2e}$ (see fig. 4). Then, with use of (8), we have

$$\left| \frac{\partial v_k}{\partial v_0} \right| = \left| \frac{\partial V_0}{\partial v}(x, v_k(x, v_0)) \right|^{-1} = \frac{|v_0|}{|v_k(x, v_0)|} \exp\left( -\int_0^{t_k(x,v_0)} \left( |v| \frac{\partial L}{\partial w} \right) (V(t', v_0)) dt' \right), \quad (17)$$

and the equation takes the form:

$$s(x) = \int_{-\infty}^{+\infty} dv_0 \sum_k N_a |v_0| \sigma_{ion}(w_k(x, v_0)) f_e(-x_c \operatorname{sgn}(v_0), v_0) +$$
$$+ \frac{m_e}{e} \int_{-\infty}^{+\infty} dv_0 \sum_k N_a |v_0| \sigma_{ion}(w_k(x, v_0)) \sum_{n: 0 < t_n(v_0) < t_k(x,v_0)} \frac{s(x_n(v_0))}{|E_n(v_0)|} \exp\left( \int_{t_n(v_0)}^0 dt' \left( |v| \frac{dL}{dw} \right) (V(t', v_0)) \right). \quad (18)$$

The number of summands on $k$ is equal to $K(x, v_0)$, in case when $K(x, v_0) = 0$ the appropriate sum is equal to zero. Here the infinity limits of integration by variable $v_0$ (also as it was by variable $v$) are formal, - really they are finite, because we have no interest for trajectories of region $\underline{A}$ (see fig. 2) and assume zero value of the distribution function here. So we can restrict this integration with limits $-v_{0m} < v_0 < v_{0m}$, $v_{0m} \equiv \max v_0 = \lim_{x \to +x_c} V_0(x, 0)$, at least, in the second term.

To make further simplifications let us rewrite (18) in the form

$$s(x) = \int_{-\infty}^{+\infty} dv_0 \sum_k N_a |v_0| \sigma_{ion}(w_k(x, v_0)) f_e(-x_c \operatorname{sgn}(v_0), v_0) +$$

$$+ \frac{m_e}{e} \sum_{n=1}^{\infty} \int_{-v_{0m}}^{+v_{0m}} dv_0 \sum_k N_a |v_0| \sigma_{ion}(w_k(x, v_0)) \theta(t_k(x, v_0) - t_n(v_0)) \frac{s(x_n(v_0))}{|E_n(v_0)|} \exp\left(-\int_0^{t_n(v_0)} dt' \left(|v| \frac{dL}{dw}\right)(V(t', v_0))\right).$$

Here the Heaviside function $\theta(z) = 0, z \leq 0;\ \theta(z) = 1, z > 0$ helps us to extend formally summation on variable $n$ into all its possible values and make it independent on the other variables.

Let $a_n = |x_n(v_{0m})|, n = 1, 2, ...$ be reverse points for the separatrix trajectory. In second integral we change the integration variable $v_0$ in every summand on $n$ to variable $x'$ on sets $(-a_n < x' < -a_{n+1}) \vee (a_{n+1} < x' < a_n)$ according to relation $v_0 = V_0(x', 0)$. Here $x' = x_n(v_0)$ are roots of this equation, - turning points of a trajectory, mentioned above. For Jacobian of transformation with the account of (9) and (14) we have:

$$\left|\frac{\partial v_0}{\partial x'}\right| = \left|\frac{\partial V_0}{\partial x}(x', 0)\right| = \frac{e}{m_e |V_0(x', 0)|} |E(x')| \exp\left(\int_0^{T(x', 0)} \left(|v| \frac{\partial L}{\partial w}\right)(V(t', V_0(x', 0))) dt'\right). \tag{19}$$

Substitution of this expression to the second integral, with the account of $T(x', 0) = t_n(V_0(x', 0))$, leads the equation to a form:

$$s(x) = \int_{-\infty}^{+\infty} dv_0 \sum_k N_a |v_0| \sigma_{ion}(w_k(x, v_0)) f_e(-x_c \operatorname{sgn}(v_0), v_0) +$$

$$+ \sum_{n=1}^{\infty} \left(\int_{-a_n}^{-a_{n+1}} dx' + \int_{a_{n+1}}^{a_n} dx'\right) s(x') \sum_k N_a \sigma_{ion}(w_k(x, V_0(x', 0))) \theta(t_k(x, V_0(x', 0)) - T(x', 0)).$$

As the expression under integrals in the second term is formally independent on $n$, we can connect neighbor intervals of integration and do a summation on $n$. Because an electron loses its total energy $\varepsilon(x, v) = -\varphi(x) + \frac{m_e v^2}{2e}$ during its motion, also due to inequality $t_k(x, V_0(x', 0)) > T(x', 0)$, which the Heaviside function provides, we have $\varepsilon(x, v_k) < \varepsilon(x', 0)$; therefore $-\varphi(x') > -\varphi(x) + \frac{m_e v_k^2}{2e} > -\varphi(x)$, and then $x_c > |x'| > |x|$, because the electric potential $\varphi(x)$ is even function, monotone decreasing in right-hand side $0 < x < x_c$. Thus, integration on variable $x'$ can be restricted by two sets: $-x_c < x' < -|x|$ and $|x| < x' < x_c$:

$$s(x) = \int_{-\infty}^{+\infty} dv_0 \sum_k N_a |v_0| \sigma_{ion}(w_k(x, v_0)) f_e(-x_c \operatorname{sgn}(v_0), v_0) +$$

$$+ \left(\int_{-x_c}^{-|x|} dx' + \int_{|x|}^{x_c} dx'\right) s(x') \sum_k N_a \sigma_{ion}(w_k(x, V_0(x', 0))) \theta(t_k(x, V_0(x', 0)) - T(x', 0)).$$

By changing a sign of integration variable $x'$ in the negative interval, also using symmetry $s(-x) = s(x)$, the equation can be transformed to the form

$$s(x) = \int_{-\infty}^{+\infty} dv_0 \sum_k N_a |v_0| \sigma_{ion}(w_k(x,v_0)) f_e(-x_c \, \text{sgn}(v_0), v_0) +$$

$$+ \int_x^{x_c} dx' s(x') \left( \sum_k N_a \sigma_{ion}(w_k(x, V_0(-x',0))) \theta(t_k(x, V_0(-x',0)) - T(-x',0)) + \right. \quad (20)$$

$$\left. + \sum_k N_a \sigma_{ion}(w_k(x, V_0(x',0))) \theta(t_k(x, V_0(x',0)) - T(x',0)) \right), \quad 0 < x < x_c. \quad \text{HCD}$$

If the anode is made with wire meshes and therefore it is transparent for ionizing electron (HCD geometry), then sums on $k$ in (20) might have several non-zero summands. It is a consequence of electron oscillations, named as a pendulum effect.

But if the anode is made with a solid metal plate (SGD geometry), it is not transparent for any electron. Then hollow cathode electron oscillations are absent. So we can consider only right-hand side of the device, where every trajectory of cathode electron has its origin at $x = +x_c$, $v < 0$, every trajectory of secondary electron has its origin in the nearest crossing of the whole trajectory solution with an abscissa axis (see fig. 5). Then integration on variable $v_0$ in the first term of (20) should be restricted with negative values only, in second term $k$ in summing with argument $-x'$ has no value, in summing with argument $x'$ it has single value $k = 1$. So the expression (20) in SGD gets a form

$$s(x) = \int_{-\infty}^{0} dv_0 |v_0| f_e(x_c, v_0) N_a \sigma_{ion}(w(x,v_0)) + \int_x^{x_c} dx' s(x') N_a \sigma_{ion}(w(x, V_0(x',0))), \; 0 < x < x_c. \; \text{SGD} \quad (21)$$

With respect to $s(x)$ we have obtained a Volterra 2-nd kind integral equation [4], (20) – for HCD geometry, or (21) – for SGD, with regular kernel, which is especially convenient for calculation. If its solution is found, the distribution function $f_e(x,v)$ can be obtained by formula (15).

Let cathode electrons be monochromatic (have a single value of initial velocity $v = -v_{ec}$ at $x = +x_c$ and $v = v_{ec}$ at $x = -x_c$):

$$f_e(x_c, v) = n_{ec} \delta(v + v_{ec}) = \frac{J_{ec}}{ev_{ec}} \delta(v + v_{ec}), \quad f_e(-x_c, v) = n_{ec} \delta(v - v_{ec}) = \frac{J_{ec}}{ev_{ec}} \delta(v - v_{ec}). \quad (22)$$

Then from (20) we obtain:

$$S(x) = \sum_k N_a \sigma_{ion}(w_k(x, -v_{ec})) + \sum_k N_a \sigma_{ion}(w_k(x, v_{ec})) +$$

$$+ \int_x^{x_c} dx' S(x') \left( \sum_k N_a \sigma_{ion}(w_k(x, V_0(-x',0))) \theta(t_k(x, V_0(-x',0)) - T(-x',0)) + \right. \quad (23)$$

$$\left. + \sum_k N_a \sigma_{ion}(w_k(x, V_0(x',0))) \theta(t_k(x, V_0(x',0)) - T(x',0)) \right), \quad 0 < x < x_c. \quad \text{HCD}$$

where $S(x) \equiv s(x) e / J_{ec}$ is a number of ionization acts per unit of path, which corresponds to one cathode electron.

For a case of simple glow discharge without hollow cathode effect, from (21) we have

$$S(x) = N_a \sigma_{ion}(w(x, -v_{ec})) + \int_x^{x_c} dx' S(x') N_a \sigma_{ion}(w(x, V_0(x',0))), \quad 0 < x < x_c. \quad \text{SGD}. \quad (24)$$

If we guess local dependence $\sigma_{ion}(x) = \sigma_{ion}(w(x))$ by neglecting a dependence on second argument, $V_0(x',0)$, in the kernel of integral in (24), the equation can be simplified further:

$$S(x) = \alpha(x)\left(1 + \int_x^{x_c} dx' S(x')\right), \quad \alpha(x) \equiv N_a \sigma_{ion}(w(x)).$$

As an electron current density is equal to $J_e(x) = J_{ec}\left(1 + \int_x^{x_c} dx' S(x')\right)$, $\frac{dJ_e}{dx} = -J_{ec} S(x)$, the equation (24) gets equivalent to well known continuity equation $\frac{dJ_e}{dx} = -\alpha(x) J_e$ from local Engel and Shtenbek cathode dark space theory. Thus, equations (23), (24) give non-local extension of classical ionization theory.

### 5. Conclusions

In this work we have derived one-dimensional Volterra 2-nd kind integral equation for source of ionization in plane hollow cathode discharge, which has an integrable kernel, - (20) or (23) – for monochromatic case. In case of simple plane glow discharge with solid anode plate when electron oscillations are impossible, the kernel of the integral equation has a simplified form (21), or (24). Solution of this equation enables to obtain kinetic distribution function for ionizing electrons, formulate a self-sustained discharge condition and compose a self-consistent problem for electric field.

In works of A. S. Metel and S. P. Nikulin [5, 6] the first attempts were made to obtain a theoretical current voltage characteristic of hollow cathode device, but they guessed small damping in energy of ionizing electron during one oscillation, so their approach was available only for low pressure when there were a lot of oscillations. Also, it was impossible to extend the ionization model over limits of pendulum motion and compare a simple glow discharge with a hollow cathode glow discharge.

The absence of non-local source model from Paschen invention of a hollow cathode in 1916 till today forced to use the Monte-Carlo methods. It meant the absence of any equation for a source of ionization in a hollow cathode!

In solving the self-consistent problem, using integral equation for ionization source derived here, it is possible to obtain current voltage characteristics in wide region of electric current densities and to make the comparison mentioned [7].

Work is made under support of Pohang Accelerator Laboratory, - Pohang, Republic of Korea, and Institute of Physics of Ukrainian National Academy of Science, - Kiev, Ukraine. Author expresses his great gratitude to Dr. A. Rudenko for invaluable help in preparing the manuscript to printing.

### Appendix 1

To obtain (8) let us find an explicit expression for reverse matrix

$$\begin{bmatrix} \frac{\partial T}{\partial x} & \frac{\partial T}{\partial v} \\ \frac{\partial V_0}{\partial x} & \frac{\partial V_0}{\partial v} \end{bmatrix} = \begin{bmatrix} \dot{x} & \frac{\partial X}{\partial v_0} \\ \dot{v} & \frac{\partial V}{\partial v_0} \end{bmatrix}^{-1} = \frac{1}{\dot{x}\frac{\partial V}{\partial v_0} - \dot{v}\frac{\partial X}{\partial v_0}} \begin{bmatrix} \frac{\partial V}{\partial v_0} & -\frac{\partial X}{\partial v_0} \\ -\dot{v} & \dot{x} \end{bmatrix}.$$

Now from the last row and column we express $\partial V_0 / \partial v$:

$$\frac{\partial V_0}{\partial v} = \frac{\dot{x}}{\dot{x}\frac{\partial V}{\partial v_0} - \dot{v}\frac{\partial X}{\partial v_0}} = \frac{1}{\frac{\partial V}{\partial v_0} - \frac{\dot{v}}{\dot{x}}\frac{\partial X}{\partial v_0}} = \left(\frac{\partial V}{\partial v_0} - \frac{\dot{v}}{v}\frac{\partial X}{\partial v_0}\right)^{-1}.$$

From equations $dx/dt = v$, $dv/dt = \dot{v}(x,v)$ we find evolution equations for partial derivatives $\partial V/\partial v_0$, $\partial X/\partial v_0$ by taking derivative $\partial/\partial v_0$ on independent variable $v_0$ and changing places of independent derivatives:

$$\begin{cases} \dfrac{d}{dt}\dfrac{\partial X}{\partial v_0} = \dfrac{\partial V}{\partial v_0}, \\ \dfrac{d}{dt}\dfrac{\partial V}{\partial v_0} = \dfrac{\partial \dot{v}}{\partial x}\dfrac{\partial X}{\partial v_0} + \dfrac{\partial \dot{v}}{\partial v}\dfrac{\partial V}{\partial v_0}. \end{cases}$$

With the use of these relations let us find an expression for the time derivative below (we designate $\ddot{v} \equiv \dfrac{d\dot{v}}{dt} = \dfrac{\partial \dot{v}}{\partial x}\dot{x} + \dfrac{\partial \dot{v}}{\partial v}\dot{v}$) and make simplifications:

$$\frac{d}{dt}\left(v\frac{\partial V}{\partial v_0} - \dot{v}\frac{\partial X}{\partial v_0}\right) = \dot{v}\frac{\partial V}{\partial v_0} + v\frac{d}{dt}\frac{\partial V}{\partial v_0} - \ddot{v}\frac{\partial X}{\partial v_0} - \dot{v}\frac{d}{dt}\frac{\partial X}{\partial v_0} =$$

$$= v\left(\frac{\partial \dot{v}}{\partial x}\frac{\partial X}{\partial v_0} + \frac{\partial \dot{v}}{\partial v}\frac{\partial V}{\partial v_0}\right) - \ddot{v}\frac{\partial X}{\partial v_0} =$$

$$= \left(v\frac{\partial \dot{v}}{\partial x} - \ddot{v}\right)\frac{\partial X}{\partial v_0} + v\frac{\partial \dot{v}}{\partial v}\frac{\partial V}{\partial v_0} = \frac{\partial \dot{v}}{\partial v}v\frac{\partial V}{\partial v_0} - \left(\frac{\ddot{v}}{\dot{v}} - \frac{v}{\dot{v}}\frac{\partial \dot{v}}{\partial x}\right)\dot{v}\frac{\partial X}{\partial v_0} =$$

$$= \frac{\partial \dot{v}}{\partial v}v\frac{\partial V}{\partial v_0} - \left(\frac{1}{\dot{v}}\left(\frac{\partial \dot{v}}{\partial x}\dot{x} + \frac{\partial \dot{v}}{\partial v}\dot{v}\right) - \frac{v}{\dot{v}}\frac{\partial \dot{v}}{\partial x}\right)\dot{v}\frac{\partial X}{\partial v_0} =$$

$$= \frac{\partial \dot{v}}{\partial v}\left(v\frac{\partial V}{\partial v_0} - \dot{v}\frac{\partial X}{\partial v_0}\right).$$

So, we have

$$\frac{d}{dt}\ln\left(v\frac{\partial V}{\partial v_0} - \dot{v}\frac{\partial X}{\partial v_0}\right) = \frac{\partial \dot{v}}{\partial v},$$

$$\ln\left(v\frac{\partial V}{\partial v_0} - \dot{v}\frac{\partial X}{\partial v_0}\right) = \ln v_0 + \int_0^t dt' \frac{\partial \dot{v}}{\partial v}(X(t',v_0),V(t',v_0)),$$

Here in defining of integration constant we account

$$V(t,v_0)\big|_{t\to +0} \approx v_0 + \dot{v}\big|_{t=0}t, \quad v\big|_{t=0} = v_0, \quad \left.\frac{\partial V}{\partial v_0}\right|_{t=0} = 1, \quad X(t,v_0)\big|_{t\to +0} \approx \pm x_c + v_0 t, \quad \left.\frac{\partial X}{\partial v_0}\right|_{t=0} = 0.$$

$$v\frac{\partial V}{\partial v_0} - \dot{v}\frac{\partial X}{\partial v_0} = v_0 \exp\left(\int_0^t dt' \frac{\partial \dot{v}}{\partial v}(X(t',v_0),V(t',v_0))\right),$$

$$\frac{\partial V}{\partial v_0} - \frac{\dot{v}}{v}\frac{\partial X}{\partial v_0} = \frac{v_0}{v}\exp\left(\int_0^t dt' \frac{\partial \dot{v}}{\partial v}(X(t',v_0),V(t',v_0))\right),$$

$$\frac{\partial V_0}{\partial v} = \left(\frac{\partial V}{\partial v_0} - \frac{\dot{v}}{v}\frac{\partial X}{\partial v_0}\right)^{-1} = \frac{v}{v_0}\exp\left(-\int_0^t dt' \frac{\partial \dot{v}}{\partial v}(X(t',v_0),V(t',v_0))\right).$$

Substituting $t = T(x,v)$, $v_0 = V_0(x,v)$, also $\dfrac{\partial \dot{v}}{\partial v}(x,v) = -|v|\dfrac{dL}{dw}$ from the equations (1), we obtain (8). Then (9) is a consequence of (6).

## Appendix 2

To solve the equation

$$v\frac{\partial f}{\partial x} + \dot{v}\frac{\partial f}{\partial v} + af = r, \quad \dot{v} \equiv -\frac{e}{m_e}\left(E(x) + \text{sgn}(v)L\left(\frac{m_e v^2}{2e}\right)\right), \quad a \equiv -|v|\frac{dL}{dw}, \quad r \equiv s(x)\delta(v)$$

with respect to function $f = f(x,v)$ first we find a solution of equation

$$v\frac{\partial f}{\partial x} + \dot{v}\frac{\partial f}{\partial v} = 0.$$

The solution of this equation is an arbitrary first integral of ODE system

$$\begin{cases} \dfrac{dx}{dt} = v, \\ \dfrac{dv}{dt} = \dot{v}(x,v). \end{cases}$$

In particular, function $v_0 = V_0(x,v)$ is a first integral of this ODE system. The system does not have more independent first integrals, so, complementary function is

$$f(x,v) = \phi(V_0(x,v)),$$

where $\phi = \phi(v_0)$ is an arbitrary function. If our aim is to express a solution inside the device area at $-x_c < x < +x_c$ through its boundary value at $x = \pm x_c$, we define $\phi(v_0) = f(\pm x_c, v_0)$. Since an electron starts from the left-hand side cathode plate with positive velocity and from the right-hand side – with negative one, we define more exactly $\phi(v_0) = f(-x_c \, \text{sgn}(v_0), v_0)$.

Now let us solve an equation

$$v\frac{\partial f}{\partial x} + \dot{v}\frac{\partial f}{\partial v} + af = 0.$$

It can be presented as

$$\frac{df}{dt} + af = 0, \quad \frac{df}{dt} \equiv \dot{x}\frac{\partial f}{\partial x} + \dot{v}\frac{\partial f}{\partial v}.$$

So, along the trajectory we have

$$\bar{f}(t) = \bar{f}_0 \exp\left(-\int_0^t \bar{a}(t')dt'\right) = \bar{f}_0 \exp\left(-\int_0^t a(X(t',v_0), V(t',v_0))dt'\right),$$

where $x = X(t,v_0)$, $v = V(t,v_0)$ is some trajectory. (Over-bar is used to distinguish dependence in function $f$ on new variables $t, v_0$.) With use of function $t = T(x,v)$ we define time here, and define constant $\bar{f}_0 = f(-x_c \, \text{sgn}(v_0), v_0)$, and $v_0 = V_0(x,v)$:

$$f(x,v) = f(-x_c \, \text{sgn}(V_0(x,v)), V_0(x,v)) \exp\left(-\int_0^{T(x,v)} a(X(t', V_0(x,v)), V(t', V_0(x,v)))dt'\right).$$

Now we return back to initial problem of this section. Making a transformation of variables $x, v \to t, v_0$ according to trajectory solutions $x = X(t,v_0)$, $v = V(t,v_0)$, we rewrite the initial problem as

$$\frac{df}{dt} + af = r.$$

Along every trajectory it is a linear non-uniform ODE. So, we can use a solution of a constant variation method:

$$\bar{f}(t) = \bar{f}_0(t) \exp\left(-\int_0^t \bar{a}(t')dt'\right),$$

$$\dot{\bar{f}}_0(t)\exp\left(-\int\limits_0^t \bar{a}(t')dt'\right) = \bar{r}(t),$$

$$\bar{f}_0(t) = \bar{f}_0(0) + \int\limits_0^t dt''\bar{r}(t'')\exp\left(\int\limits_0^{t''}\bar{a}(t')dt'\right),$$

$$\bar{f}(t) = \bar{f}_0(0)\exp\left(-\int\limits_0^t \bar{a}(t')dt'\right) + \int\limits_0^t dt''\bar{r}(t'')\exp\left(\int\limits_t^{t''}\bar{a}(t')dt'\right).$$

Substituting

$$\bar{f}(t) = \bar{f}(t, v_0) = f(x, v),$$

$$\bar{f}_0(0) = f(-x_c \operatorname{sgn}(v_0), v_0), \quad \bar{a}(t') = a(X(t', v_0), V(t', v_0)), \quad \bar{r}(t'') = r(X(t'', v_0), V(t'', v_0))$$

then $t = T(x, v)$, $v_0 = V_0(x, v)$, with an account to previous definitions for functions $a, r$, we obtain

$$f(x,v) = f(-x_c \operatorname{sgn}(V_0(x,v)), V_0(x,v))\exp\left(\int\limits_0^{T(x,v)} dt'\left(|v|\frac{dL}{dw}\right)(V(t', V_0(x,v)))\right) +$$

$$+ \int\limits_0^{T(x,v)} dt'' s(X(t'', V_0(x,v)))\delta(V(t'', V_0(x,v)))\exp\left(\int\limits_{t''}^{T(x,v)} dt'\left(|v|\frac{dL}{dw}\right)(V(t', V_0(x,v)))\right).$$

## Bibliography

1. A. Engel, M. Shtenbek. *Physics and technology of electric discharge in gases*, **2**, 1936.
2. Yu. M. Kagan. J. Phys. D: Appl. Phys., **18**, 1113 (1985).
3. Р. Курант. *Уравнения в частных производных*, (Мир, Москва, 1964).
4. В. Вольтерра, *Теория функционалов, интегральных и интегро-дифференциальных уравнений*, (Наука, Москва, 1982).
5. А. С. Метель, ЖТФ, **55**, 10, 1928 (1985).
6. С. П. Никулин, ЖТФ, **62**, 12, 21 (1992).
7. V. V. Gorin. Ukrainian J. Phys., **53**, 4, 366 (2008)
 http://www.ujp.bitp.kiev.ua/papers/530411p.pdf

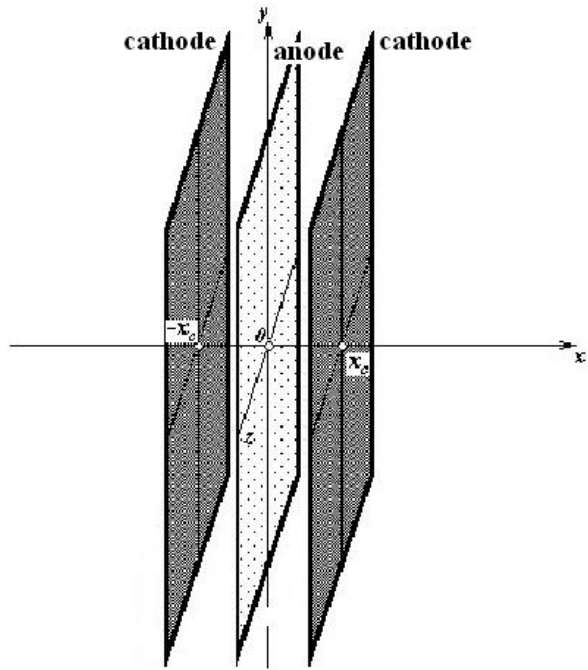

***Fig. 1.*** *Plane hollow cathode configuration.*

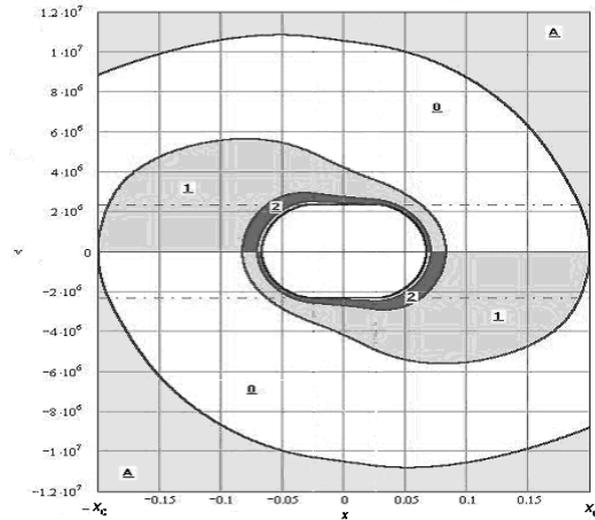

***Fig. 2.*** *Phase plane, divided on the regions having a definite number of secondary electrons. A secondary electron assumed as originated in abscissa axis in discharge space at v = 0. "A" is a region for trajectories of very fast cathode electrons, which are passing through the whole discharge space and reach another cathode; we guess such electrons are to be absent: $f_e(x,v)=0, (x,v)\in \underline{A}$. "0" is a region for cathode electron trajectories only. "1", "2" and so on are the regions, where a trajectory of cathode electron and one, two, and so on, trajectories of secondary electron are coincident. So, the longest is a cathode electron trajectory, trajectory of secondary electron is some piece of appropriate cathode electron trajectory.*

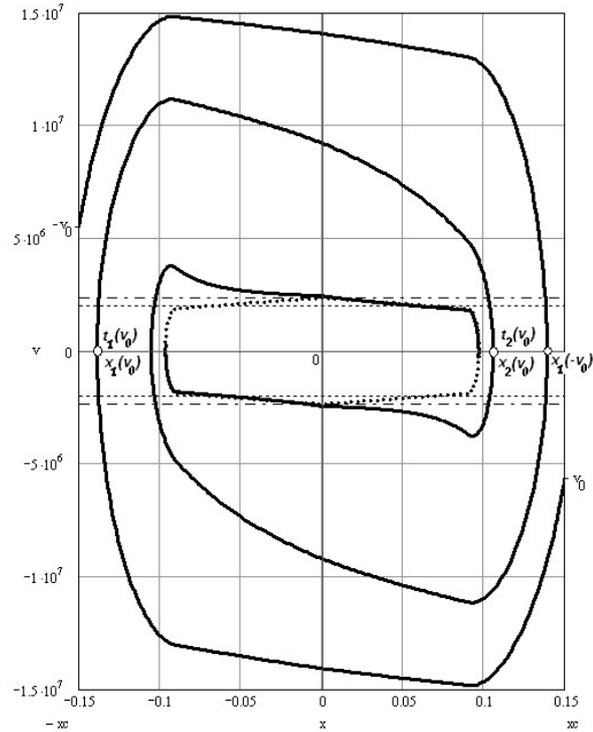

***Fig. 3.*** *Two symmetrical phase trajectories of cathode electron in HC geometry.*
*Here the anode position is $x = 0$, the cathode position is $x = \pm 0.15$, dotted horizontal lines correspond to excitation energy with velocity $v = \pm v_{ex}$, dashed horizontal lines correspond to ionization energy with velocity $v = \pm v_{ion}$. The reverse points of electron motion for instant times $t_n(v_0)$, $n = 1, 2$ are shown in crossing a trajectory and abscissa axis.*

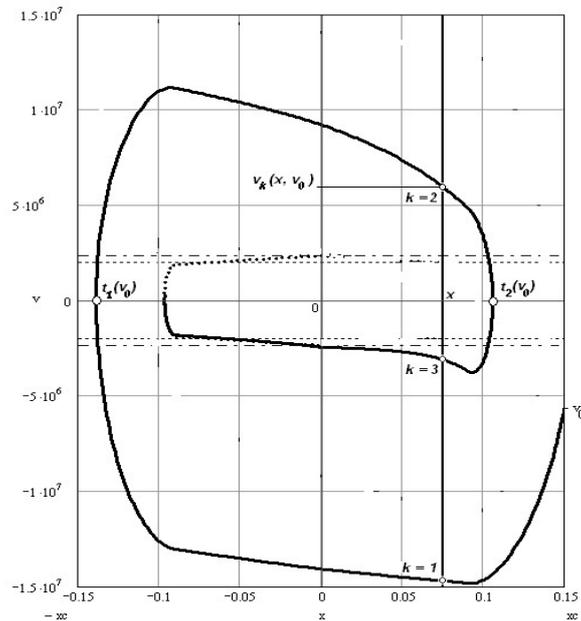

***Fig. 4.*** *In pendulum oscillations in HCD geometry an electron can reach the same anode distance for several times. So, one value of initial velocity $v_0$ in the cathode corresponds to several (here – three) values of velocity $v$ at positive coordinate $x$, because an equation $X(t, v_0) = x$ can have several roots $t = t_k$, $k = 1, 2,...$*

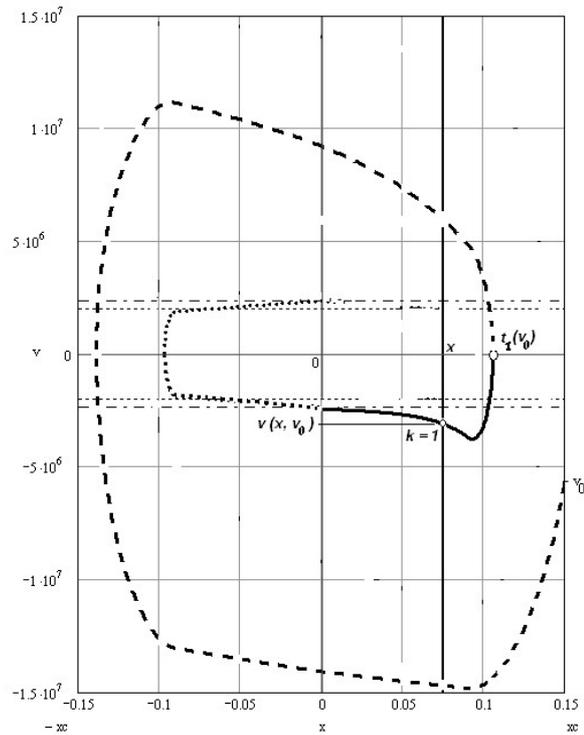

***Fig. 5.*** *In SGD geometry the anode is not transparent for any electron, and oscillations, which were typical for hollow cathode, are impossible. No secondary electron can be on the dotted piece of the trajectory solution. In SGD the equation $X(t, v_0) = x$ has single solution for the time variable along the piece of possible motion (solid), also a set of $v_k(x, v_0)$, $k = 1,...$ has single (or no) element $v_1 = v_1(x, v_0)$, the reverse point $t_n = t_1$ is a single – it is a start point of secondary electron.*